\documentclass[aps,prb,twocolumn,floatfix]{revtex4}
\usepackage{graphicx,amsfonts}
\usepackage{amsmath,amscd,amsfonts,amssymb,color}
\newcommand{\bra}[1]{\left\langle{#1}\right\vert}
\newcommand{\ket}[1]{\left\vert{#1}\right\rangle}
\begin{document}
\title{Quantum state estimation using weak
measurements}
\author{Debmalya Das}
\affiliation{Department of Physical Sciences, Indian Institute of
Science Education
\& Research (IISER) Mohali, Sector-81, SAS Nagar, Manauli P.O.
140306, Punjab, India.}
\email{debmalya@iisermohali.ac.in}
\author{Arvind}
\affiliation{Department of Physical Sciences, Indian Institute of
Science Education
\& Research (IISER) Mohali, Sector-81, SAS Nagar, Manauli P.O.
140306, Punjab, India.}
\email{arvind@iisermohali.ac.in}
\begin{abstract}
We explore the possibility of using ``weak
measurements'' without ``weak value'' for quantum
state estimation. Since for weak measurements the
disturbance caused during each measurement is
small, we can rescue and recycle the state, unlike for the case
of projective measurements.  We use this property
of weak measurements and design  schemes for
quantum state estimation for qubits and for
Gaussian states.  We show, via numerical
simulations, that under certain circumstances, our
method can outperform the estimation by projective
measurements.
It turns out that ensemble size plays an
important role and the scheme based on recycling
works better for small ensembles.
\end{abstract}
\maketitle
\section{Introduction}
The quantum superposition principle and the
wave function collapse set apart the quantum
description of the world from its classical
counterpart. As a consequence, in the standard
paradigm, the outcome of a single measurement
cannot be predicted with  certainty and we can
only assign probabilities to different outcomes.
Since the measurement process disturbs the system
and in a projective measurement the state of the
system collapses we cannot re-use the state 
for any further measurement.  This necessitates
the use of an ensemble of identically prepared
states in order to interpret quantum measurements.
Ideally, in the large ensemble limit the ensemble
average tends to the expectation value of the
observable.  The question that arises at this
point is: what if we are provided with a small
ensemble of states and asked to make the best use
of it?

Projective measurements require a large coupling
between the system and the measuring device.
However, if the coupling is made small, we inflict
a very small disturbance to the system at the
expense of extracting a correspondingly small amount of
information~\cite{brun}.  Such measurements are
known as weak or unsharp measurements. Such
measurements have been  introduced in various
forms in the
past~\cite{Busch,kraus1983,AliFuzzy,Prugovecki1,Prugovecki2,
Diosi,AAV}. The coupling strength can be tuned to suit
the situation and the state can subjected to
further measurements to extract more information.
Whenever we have a small ensemble, each
member can be ``weakly'' measured more than once
with a possibility of extracting  more
information.

It is that true that all quantum measurements (projective,
non-projective, weak etc) can be seen as Positive
Operator Valued Measures (POVM). Still it is important
to know the details and workings of a measurement scheme. 
A
POVM can also be interpreted as a projective
measurement on a larger Hilbert
space~\cite{NC,P3,brun}.  For a finite ensemble
the upper bound on the amount of information
extractable is available~\cite{MassPop1}. There is
always a  cost
of  information extraction from quantum systems in
terms of the disturbance caused and that too has 
has also been explored
for the case of weak measurements
~\cite{Ueda,Branciard,Cheong,dass}.

A recent work by Rozema et. al. suggests some new
possibilities that weak measurements can offer with respect to
Heisenberg's uncertainty relation and the disturbance caused
to the state~\cite{distwm}.  Oreshkov et. al., in 2005,
wrote down a weak measurement POVM and showed that any
generalized measurement can be decomposed into a sequence of
weak measurements, without using an ancilla~\cite{wmPOVM}.
Lundeen et. al. recently came up with a method employing
weak values to
directly measure the wave function of a quantum system in a
pure state~\cite{Qwavefn} and followed it up with a method
to measure any general state~\cite{GenQwavefn}.
For some further developments in this
regard see~\cite{phyQwavefn}. Unsharp measurements have also
been used to make sequential measurements on a single
qubit~\cite{Diosi}. Other examples of quantum
state tomography with weak measurements can be found in
~\cite{sttomowm,sswm,Kob2014}. An approach to perform quantum state
tomography using weak measurement POVMs was introduced by
Hofmann~\cite{Hofmann}.

We present in
this paper some of our results on state estimation
by ``weak'' measurements\footnote{A more detailed account
for the qubit case is available in our recent
paper~\cite{Qsize}}. We 
explore the case of a single qubit and
show by
explicit simulations how under certain
circumstances the weak measurement based state
estimation scheme can beat the one based on
projective measurements. 
\section{Weak and unsharp measurements}
\label{weak-measurement}
The measuring apparatus plays a crucial role in
quantum measurements; on the one hand it interacts
with the quantum system and on the other hand it
has classical properties where the outcomes can be
read out and recorded.  A useful model of this
process is available due to von Neumann. Although
originally this model was constructed for strong
(projective) measurements~\cite{johnV} it has
wider applications and can also be applied to weak
measurements~\cite{Busch,kraus1983,AliFuzzy,Prugovecki1,Prugovecki2,
Diosi,AAV,Sudweak}.

\subsection{Von Neumann's measurement model for discrete basis}
Consider the measurement of an observable $A$ of a quantum
system with eigenvectors $\{\vert a_j\rangle\}$ and
eigenvalues $\{a_j\}$, $j=1\cdots n$. Imagine an apparatus
with continuous pointer positions described by a variable
$q$ and its conjugate variable $p$ such that $[q,p]=i$.
The initial state of the measuring device has an initial
spread of $\Delta q$ with its Gaussian quantum state
$\ket{\phi_{in}}$ centered around zero given by 
\begin{equation}
\ket{\phi_{in}}=\left(\frac{\kappa}{2\pi}\right)^{\frac{1}{4}}\int_{-\infty}
^\infty dq \,e^{-\frac{\kappa q^2}{4}}\ket{q}
\label{pointer_in}
\end{equation}
where $\kappa=\frac{1}{\left(\Delta q\right)^2}$ and we
have taken $\hbar=1$.  The system and the measuring device
are made to interact by means of a Hamiltonian,
\begin{equation}
\label{hamiltonian}
 H=g\delta\left(t-t^\prime\right)A\otimes p
\end{equation}
where $p$ is the momentum conjugate to the variable $q$,
and $g$ is the coupling strength. The
Hamiltonian is so chosen that the system and the device get
a kick and interact momentarily at $t=t^\prime$. Let the
initial state $\ket{\psi_{in}}$ of the system be written in
terms of the eigenstates $\ket{a_1},
\ket{a_2},.....,\ket{a_n}$ of the operator $A$.
\begin{equation}
 \ket{\psi_{in}}=\sum_{i=1}^n c_i \ket{a_i}
\end{equation}
The joint evolution of the system and the measuring device
under the coupling Hamiltonian gives an entangled
state for $t> t^{\prime}$
\begin{eqnarray}
&&e^{-i \int H
dt}\ket{\psi_{in}}\otimes\ket{\phi_{in}}=\nonumber \\
&&\left(\frac{\kappa }{2\pi}\right)^{\frac{1}{4}}\sum_{i=1}^n
\int_{-\infty}^\infty dq c_i \,e^{-\frac{\kappa \left(q-ga_i\right)^2}{4}}\ket{a_i}\otimes\ket{q}
\label{grand}
\end{eqnarray}
The above state consists of a series of Gaussians centered
at $g a_1, g a_2,\cdots,g a_n$ for the pointer entangled
with corresponding eigenstates $\vert a_1\rangle,\vert
a_2\rangle \cdots \vert a_n\rangle$ of the system.  At this
stage we invoke the fact that the apparatus is
classical, because of
the fact that only one of the pointer positions actually
shows up. This requires the collapse of the wave function
which is brought in as something from outside for the classical
apparatus!  Thus the  process is completed with the meter
showing only one of the $g a_i$s and  the system
state collapses into the corresponding eigenstate $\vert
a_i\rangle$.  The above analysis holds good only if the
Gaussians are well separated or distinct. In
contrast, when the Gaussians
overlap, which can happen if the coupling strength $g$ is
small or the initial spread in the pointer state given by
$1/\kappa$  is large, the scenario changes~\cite{Busch,AAV,Marc}.
This is called the weak or unsharp measurement regime.  
Weak measurements
have been employed in developing recipes for the violation
of Bell inequalities~\cite{Marc} and Leggett-Garg
inequalities~\cite{LGineq}.  These have also been recently
used to study super-quantum discord~\cite{patisingh3}. 
\subsection{Weak values and post-selection}
In the treatment of weak measurement given by
Aharonov, Albert and Vaidmann (AAV), first a
subsequent projective measurement of a second
observable $B$ is carried out, followed by a
post-selection of the output state into one of the
eigenstates of the second observable, say
$\ket{b_j}$.  The weak value of the observable
$A$, which was measured in the weak regime, is then defined as
\begin{equation}
A_w=\frac{\bra{b_j}A\ket{\psi_{in}}}{\bra{b_j}\psi_{in}\rangle}
\label{weakvalue} \end{equation} When the
post-selected state $b_j$ is nearly orthogonal to
the initial state $\ket{\psi_{in}}$
equation[~\ref{weakvalue}] tells us that the weak
value becomes  very large, so large that it
can lie outside the allowed range of the
eigenvalue spectrum ~\cite{AAV,Sudweak}.

The interpretation of weak values is a current
topic of research in quantum information
theory. Weak values can be complex and the real
and the complex parts can be interpreted in terms
of the displacements in the position and momentum
spaces, respectively, of the measuring device
~\cite{Joz}. Weak values have been used to reinterpret
the flow of time in quantum mechanics
~\cite{Toll} and in the direct measurement of the
photon wavefunction ~\cite{Qwavefn} and in the
amplification of small signals
~\cite{ampinterferometer, ampnoise, amplimit}.
Another interesting application of weak values is
in connection with quantum Chesire cat
experiments~\cite{Chesire, matzkin}. There has
been criticisms of the method of post-selection as
well, namely that the process
of post-selection leads to throwing away data and
can lead to suboptimal use of information from a
measurement. For discussions on the same see
~\cite{Combes,Ferrie,Vaidman2014}. However, we take
a different approach in our work, where we do not
do any post-selection i.e. we consider {\em weak
measurements without weak values}.
\subsection{Effect of weak and strong measurement on a qubit}
How exactly do we carry out the weak measurement?
How much is the effect of a weak measurement on
the system?  If we carry out weak measurements on
all the members of an identically prepared
ensemble, what happens to such an ensemble? We
illustrate these points by taking an example.
Consider a measurement of $\sigma_z$ ($z$
component of spin) of a qubit in a fixed quantum
state. Following the general prescription given in
Equation~(\ref{hamiltonian}) we write the
interaction Hamiltonian as 
\begin{equation}
\label{interaction}
 H=g\delta\left(t-t^\prime\right)\sigma_z\otimes p
\end{equation}
assuming the initial state of the pointer to be
the same as that
given in Equation~(\ref{pointer_in}).  The qubit
is taken to be 
in a pure
state given by 
\begin{equation}
\vert \psi_{in} \rangle = \cos{\frac{\alpha}{2}} \vert 0
\rangle+ \sin{\frac{\alpha}{2}}\vert 1 \rangle
\end{equation}
where $\vert 0 \rangle$ and $\vert 1 \rangle$ are the
eigenstates of $\sigma_z$ with eigen values $+1$ and $-1$
respectively.
The combined state of
the system and the pointer after the interaction is given by 
taking a special case of Equation~(\ref{grand}) 
\begin{eqnarray}
\vert \psi_{out}\rangle =&&
\left(\frac{\kappa }{2\pi}\right)^{\frac{1}{4}}
\int_{-\infty}^\infty dq \cos{\frac{\alpha}{2}} 
\,e^{-\frac{\kappa
\left(q-g\right)^2}{4}}\ket{0}\otimes\ket{q}\nonumber \\
&&+\left(\frac{\kappa }{2\pi}\right)^{\frac{1}{4}}
\int_{-\infty}^\infty dq \sin{\frac{\alpha}{2}} 
\,e^{-\frac{\kappa 
\left(q+g\right)^2}{4}}\ket{1}\otimes\ket{q}\nonumber \\
\label{grand_1}
\end{eqnarray}
At this stage the apparatus and the system are in an
entangled state.
An observation of the apparatus will lead to values
whose distribution is determined by the above state. It is
clear from Equation~(\ref{grand_1}) that the distribution of
values of the apparatus is a Gaussian centered around $+g$
for the system input state $\vert 0 \rangle$ and is a
Gaussian centered around $-g$ 
for the system input state $\vert 1 \rangle$.
The width of the Gaussian in  each case is given by
$1/{\kappa}$. By tuning the parameter $\epsilon=\kappa g $
we can change the nature of the measurement in terms of its
strength. In our
work we have taken $g=1$ so that we have $\epsilon=\kappa$.
For large
values of $\epsilon$ we have a projective measurement, where
the pointer distributions are well separated for the states 
$\vert 0 
\rangle$ and $\vert 1 \rangle$. Therefore, each reading of
the pointer tells us exactly what the state of the system is
after the measurement. By repeatedly measuring the same
observable we can calculate the expectation value of the
observable. The state collapses completely in each
measurement
and there is no question of re-using these states.
However, when the value of $\epsilon$ is small we have two
Gaussians that overlap. From an observation of
the pointer we do not learn with certainty as to what value
to assign to the system spin $z$ component. The pointer
positions are weakly correlated with the eigenstates of
$\sigma_z$. The state is only partially affected and there
is a possibility of re-using the state.  
The effect of the weak measurement in this
case can be explicitly calculated and it turns out that there is
very little change in the state of the system.
The final state of the system can be calculated
by taking the state in Equation~(\ref{grand}) and then
taking a partial trace over
the apparatus's degrees of freedom giving us the final  mixed state
corresponding to the system alone:
\begin{equation}
\rho_f=\frac{1}{2}\begin{pmatrix}
1+\cos{\alpha} &
(1-\frac{\epsilon}{8})\sin{\alpha}\\
(1-\frac{\epsilon}{8})\sin{\alpha} &
1-\cos{\alpha}
                    \end{pmatrix}
\end{equation}
Since $\epsilon$ is small we can conclude that the
disturbance caused to the system is also small.
Furthermore,
the disturbance can be controlled by changing $\epsilon$. 
\section{Quantum State Estimation of a single qubit}
\label{procedure}
We now turn to the question of using weak measurements with
state recycling for the problem of state estimation of a
single qubit. 

\subsection{The scheme}
\begin{figure}[h]
\includegraphics{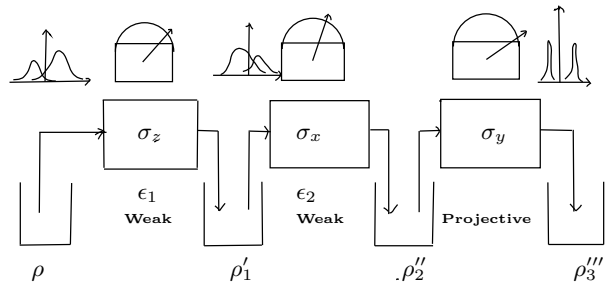}
\caption{
The schematic diagram of our prescription involving two
weak measurements of coupling strengths $\epsilon_1$
and $\epsilon_2$, allowing state recycling, followed by a
projective measurement.}
\label{flowchart_1}
\end{figure}
In our prescription, we consider a finite
size ensemble of pure
or mixed states of a qubit.
On every member of the ensemble we
carry out a $\sigma_z$ measurement whose strength is defined
by the parameter $\epsilon_1$.  We record the meter reading
in each case and keep the modified states after measurements to
obtain a changed ensemble. This new ensemble is now used to
measure $\sigma_x$ in the same way but with a coupling
strength $\epsilon_2$. Finally the resultant ensemble is
used to carry out  projective measurement of $\sigma_y$ on
its members. The first two measurements are weak while the
last measurement is strong or projective. 
To avoid statistical errors the results are  averaged 
over many runs. 
The entire process is summarized in
Figure~\ref{flowchart_1}.
For both the weak measurements,
consider a regime in which $\epsilon$ is neither too large
to make the measurement projective, nor too small, as is
done in traditional weak measurements. For such values of
$\epsilon$, the two Gaussians, representing the pointer value
distributions for the two eigen values of the observable,
overlap partially with each
other. When there is no overlap, a meter reading
unambiguously indicates an outcome and we have a projective
measurement.  A meter reading corresponding to a point in the
overlap region cannot be reliably correlated with the system
being in one or the other eigenstate.
To reduce this difficulty, let us define a
region, midway between the centers of the two Gaussians, of
width $2a$. We call it the discard region, which means that
any pointer reading which falls in this region is rejected.
For the case where we measure $\sigma_z$, 
all readings where the pointer position is to the right of this
region are interpreted as indicating the value of $\sigma_z$
to be  $+1$ while the ones on the left of this region are
interpreted as $-1$. Even when the outcome is discarded, the
member of the ensemble is not rejected, but is
retained
to be re-used for the next measurement.
In summary, in this scheme as is shown in
Figure~\ref{flowchart_1} we
first measure $\sigma_z$ weakly, followed by 
$\sigma_x$
which is again measured weakly and last we make
a projective
measurement of $\sigma_y$. The entire simulation is run on 
identically prepared copies (ensemble size) 
of the state of interest (pure or mixed). The simulation is
repeated many times to avoid statistical errors.

\par A general single qubit state is given by
\begin{equation}
\rho=\rho_{00}\ket{0}\bra{0}+
\rho_{01}\ket{0}\bra{1}+\rho_{10}\ket{1}\bra{0}
 +\rho_{11}\ket{1}\bra{1}
\end{equation}
The diagonal elements are known as populations as
they give the probabilities with which the states $\ket{0}$
and $\ket{1}$ are present in the mixture. The off-diagonal
elements are known as coherences as these contain the
phase information of the states $\ket{0}$ and $\ket{1}$.
When the state is coupled to a measurement device,
as discussed above,
the resultant state
after unitary evolution for a strength $\epsilon$, is
\begin{eqnarray}
&&\rho'=\left(\frac{\epsilon}{2\pi}\right)^\frac{1}{2} 
\nonumber \\
&&\left[\int_{-\infty}^\infty dq
\int_{-\infty}^\infty
dq'\rho_{00}\,e^{-\frac{\epsilon\left(q-1\right)^2}{4}}
\,e^{-\frac{\epsilon\left(q'-1\right)^2}{4}}\ket{0}\bra{0}+ 
\right. \nonumber\\
&&
\int_{-\infty}^\infty dq \int_{-\infty}^\infty
dq'\rho_{01}\,e^{-\frac{\epsilon\left(q-1\right)^2}{4}}
\,e^{-\frac{\epsilon\left(q'+1\right)^2}{4}}\ket{0}\bra{1}+
\nonumber\\
&&\int_{-\infty}^\infty dq \int_{-\infty}^\infty
dq'\rho_{10}\,e^{-\frac{\epsilon\left(q+1\right)^2}{4}}
\,e^{-\frac{\epsilon\left(q'-1\right)^2}{4}}\ket{1}\bra{0}+
\nonumber\\
&& \left.
\int_{-\infty}^\infty dq \int_{-\infty}^\infty
dq'\rho_{11}\,e^{-\frac{\epsilon\left(q+1\right)^2}{4}}
\,e^{-\frac{\epsilon\left(q'+1\right)^2}{4}}\ket{1}\bra{1}\right]  
\nonumber\\
&&\otimes \ket{q}\bra{q'}
\label{grand_2}
\end{eqnarray}
Let us consider taking out a member of the ensemble of
system states and then coupling it with the apparatus.
Now when the observer
notes down the meter reading,  she can see a particular
reading which depends upon the initial states of the system
and the meter and the coupling between the two. Though this
process is not well understood and von Neumann's model is
silent about this final step of collapse, it can be thought
of as the action of the projector $\ket{q}\bra{q}$ on the
meter state resulting in the meter reading $q$.

\par The probability density of obtaining the value $q$ for the meter
is therefore given by
\begin{equation}
 P(q)=Tr\left(\ket{q}\bra{q}\rho_{MD}\right)
\end{equation}
where the reduced density operator for the apparatus or the
measuring device (MD)
is obtained by taking a partial trace of the state
$\rho^\prime$ over the system. 
\begin{equation}
 \rho_{MD}=Tr_{system}\left(\rho'\right)
\end{equation}
This probability density can now be used to calculate the
probabilities of possible outcomes. For example,
$P(\sigma_z=1)$ can be
obtained by integrating the probability density from $+a$ to $\infty$.
Thus, the
probabilities with which we obtain $+1$ , $-1$ or ambiguous
readings while measuring in the $z$-basis are
calculated by integrating the above probability densities
from $+a$ to
$\infty$, $-\infty$ to $-a$ and $-a$ to $+a$, respectively
and are given by
\begin{widetext}
\begin{eqnarray}
P(\ket{0})&=&\frac{1}{4}\left[\left(1+z\right)
Erfc{\frac{\left(-1+a\right)\sqrt{\epsilon_1}}{\sqrt{2}}}-\left(-1+z\right)
Erfc{\frac{\left(1+a\right)\sqrt{\epsilon_1}}{\sqrt{2}}}\right]
\nonumber\\
P(\ket{1})&=&\frac{1}{4}\left[-\left(-1+z\right)
Erfc{\frac{\left(-1+a\right)\sqrt{\epsilon_1}}{\sqrt{2}}}+\left(1+z\right)
Erfc{\frac{(1+a)\sqrt{\epsilon_1}}{\sqrt{2}}}\right]
\nonumber\\
P(discard_z)&=&\frac{1}{2}\left[Erf{\frac{\left(-1+a\right)
\sqrt{\epsilon_1}}{\sqrt{2}}}+Erf{\frac{\left(1+a\right)
\sqrt{\epsilon_1}}{\sqrt{2}}}\right]
\end{eqnarray}
\end{widetext}
Further for the second weak measurement, the input
state is the output from the first measurement described by
an ensemble $\rho_1^\prime$. This ensemble is obtained from
the state $\rho^\prime$ given in Equation~(\ref{grand_2}) by
taking a trace over the measuring device (apparatus)
\begin{equation}
\rho^\prime_1= {\rm Tr}_{MD}(\rho^\prime)
\end{equation}

The probabilities with which we obtain the value $+1$, 
$-1$  or
ambiguous readings while measuring in the $\sigma_x$-basis are 
given by, 
\begin{widetext}
 \begin{eqnarray}
P(\ket{\sigma_x;+})&=&\frac{1}{4}e^{-\frac{\epsilon_1}{2}}
\left[\left(-Erf{\left(-1+a\right)
\sqrt{\frac{\epsilon_2}{2}}}+
Erf{\left(1+a\right)\sqrt{\frac{\epsilon_2}{2}}}\right)x+
e^{\frac{\epsilon_1}{2}}\left(
Erfc{\left(-1+a\right)\sqrt{\frac{\epsilon_2}{2}}}+
Erfc{\left(1+a\right)
\sqrt{\frac{\epsilon_2}{2}}}\right)\right]\nonumber\\
P(\ket{\sigma_x;-})&=&\frac{1}{4}e^{-\frac{\epsilon_1}{2}}
\left[\left(Erf{\left(-1+a\right)
\sqrt{\frac{\epsilon_2}{2}}}-
Erf{\left(1+a\right)\sqrt{\frac{\epsilon_2}{2}}}\right)x+
e^{\frac{\epsilon_1}{2}}
\left(Erfc{\left(-1+a\right)
\sqrt{\frac{\epsilon_2}{2}}}+Erfc{\left(1+a\right)
\sqrt{\frac{\epsilon_2}{2}}}\right)\right]\nonumber\\
P(discard_x)&=&\frac{1}{2}
\left[Erf{\frac{\left(-1+a\right)\sqrt{\epsilon_1}}
{\sqrt{2}}}+Erf{\frac{\left(1+a\right)
\sqrt{\epsilon_1}}{\sqrt{2}}}\right]
\end{eqnarray}
\end{widetext}
After this measurement if we trace over the second apparatus
we obtain the ensemble represented through a density
operator $\rho^{\prime\prime}_2$. Lastly we perform a
regular strong (projective) measurement of $\sigma_y$ and
the  probabilities are given by,
\begin{eqnarray}
P(\ket{\sigma_y;+})&=&\frac{1}{2}
\left[1+e^{-\frac{1}{2}\left(\epsilon_1+\epsilon_2\right)}y\right]
\nonumber\\
P(\ket{\sigma_y;-})&=&\frac{1}{2}\left[1-e^{-\frac{1}{2}
\left(\epsilon_1+\epsilon_2\right)}y\right]
\end{eqnarray}
In the above equations, we have used
\begin{eqnarray}
Erf(x)&=&\frac{2}{\sqrt{\pi}}\int_0 ^x e^{-t^2} dt
\nonumber\\
Erfc(x)&=&1-Erf(x)
\end{eqnarray}

These measurements when repeated over the entire ensemble
give us an estimate of the expectation values of
$\sigma_x$, $\sigma_y$ and $\sigma_z$, which in turn help us
locate the 
co-ordinates $(x,y,z)$ of the point inside the
Bloch sphere:
\begin{eqnarray}
z &=& Tr\left(\rho \sigma_z\right)
\nonumber\\
x &=& Tr\left(\rho'_1 \sigma_x\right)e^{\frac{\epsilon_1}{2}}
\nonumber\\
y &=& Tr\left(\rho''_2 \sigma_y\right)
e^{\frac{1}{2}\left(\epsilon_1+\epsilon_2\right)}
\label{estimate}
\end{eqnarray}
where $\rho$, $\rho'_1$ and $\rho''_2$ denote the initial
state of the system and those after the first and second
measurements respectively. We note that $\epsilon_1$ and
$\epsilon_2$ appear in Equation~(\ref{estimate}) 
because we are interested in the expectation
values of $\sigma_x$, $\sigma_y$ and $\sigma_z$ for the
original state $\rho$ of the system. These results are valid
only for small values of $\epsilon_1$ and $\epsilon_2$.  In
subsequent studies we work with the simplification
$\epsilon_1=\epsilon_2 =\epsilon$.

For a scheme based purely on projective measurements, the
ensemble is divided into three equal parts and direct
measurements of $\sigma_x$, $\sigma_y$ and $\sigma_z$ are
performed independently. This leads to a direct estimate of
the expectation values of these operators giving the values
of  $(x,y,z)$ and hence an estimate of the state. The error
in these estimates depends upon the size of the ensemble.
We simulate both these schemes and compare the performance
of our method with the one based on projective measurements.

We recall that a qubit can be represented as a point in a
Bloch sphere~\cite{NC,sakurai}.The Bloch sphere is a unit sphere
the pure and mixed states of a qubit lying on the surface and 
inside the sphere, respectively. The state corresponding to the
point $(x,y,z)$ is given by
\begin{equation}
 \rho=\frac{1}{2}\left(I+\vec{n}.\vec{\sigma}\right)
\end{equation}
where $\hat{n}=x\hat{x}+y\hat{y}+z\hat{z}$  is a  vector with 
$x=\langle \sigma_x\rangle$, $y=\langle \sigma_y\rangle$ and 
$z=\langle \sigma_z\rangle$. The pure states correspond to
the case when the point lies on the surface and in that case
$\vec{n}$ is a unit vector. 
The expectation values of $\sigma_x$, $\sigma_y$ and
$\sigma_z$ serve as a direct means to calculate the values of
$(x,y,z)$. Therefore, to carry out state estimation of a
given state of a
single qubit, we need to estimate
the numbers $(x,y,z)$.
The performance of our scheme is quantified using the
fidelity measure:
\begin{equation}
\label{fidelity}
 f=1-\left[\left(x-x_{est}\right)^2 +
\left(y-y_{est}\right)^2 + \left(z-z_{est}\right)^2\right]
\end{equation}
\subsection{Average performance over Bloch sphere}
We move on to test our
scheme on a large number of randomly generated states of a
qubit and look for the average performance of the scheme
over the Bloch sphere.  The  process is carried out for 2000
states generated randomly. We also study the dependence on
ensemble size and use ensemble sizes of $30$ and $60$.
For each case the simulation is repeated $1000$ times to 
average over statistical fluctuations.

\begin{figure}
\includegraphics{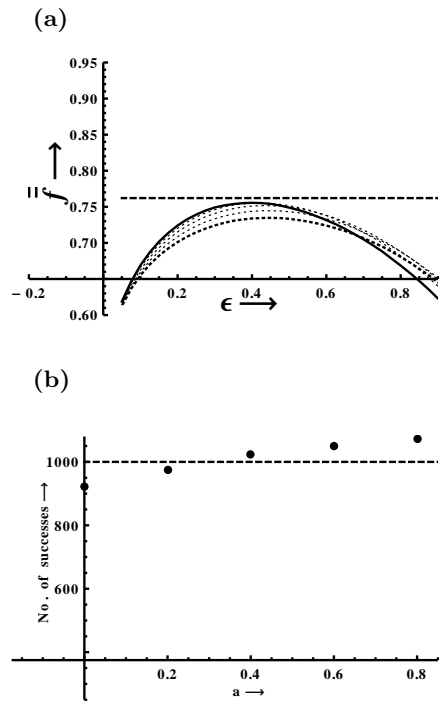}
\caption{(a) Plot of  the mean fidelity $\bar{\bar{f}}$  for
a state with ensemble size $30$ and mean calculated over
1000 runs, further averaged over 2000 randomly chosen
states, as a function of the coupling strength $\epsilon$.
Different curves represent different values of the discard
parameter $a$.  The discard parameter used are  $a=0$
(dotted thick line), $a=0.2$ (dotted line), $a=0.4$ (dotted
line) $a=0.6$
(dotted line) and $a=0.8$ (solid line). The straight dotted line
represents projective measurements. The solid line
comes very close to the projective measurements. (b) Plot of
the number of times our schemes outperform the projective
measurement based scheme for the 2000 randomly chosen states
of the qubit as a function of the discard parameter $a$.}  
\label{fid_score_30}
\end{figure}

\begin{figure}
\includegraphics{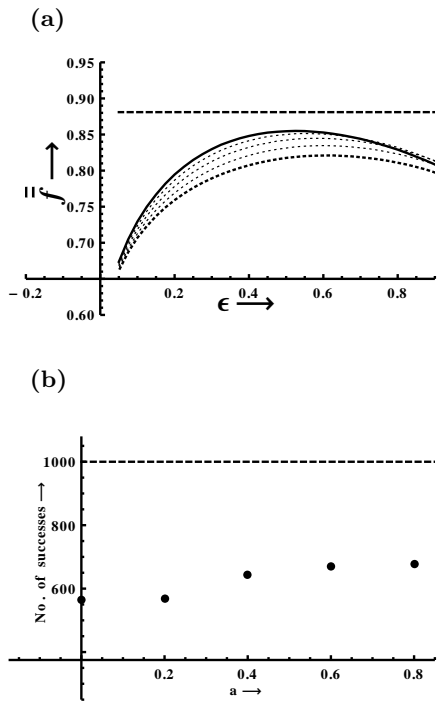}
\caption{(a) Plot of  the mean fidelity $\bar{\bar{f}}$  for
a state with ensemble size $60$ and mean calculated over
1000 runs, further averaged over 2000 randomly chosen
states, as a function of the coupling strength $\epsilon$.
Different curves represent different values of the discard
parameter $a$.  The discard parameter used are  $a=0$
(dotted thick line), $a=0.2$ (dotted line), $a=0.4$ (dotted
line) $a=0.6$
(dotted line) and $a=0.8$ (solid line). The straight dotted line
represents the  projective measurements. 
(b) Plot of
the number of times our schemes outperform the projective
measurement based scheme for the 2000 randomly chosen states
of the qubit as a function of the discard parameter $a$. 
The success rate goes down with an increase in ensemble
size from $30$ to $60$.
}
\label{fid_score_60}
\end{figure}

While we average the fidelity over all states to obtain the
average fidelity we also keep track whether the scheme
outperformed or underperformed as compared to the
projective measurement scheme in each case.  For the ensemble size of
$30$, the results of this simulation are presented in two
different ways in Figure~\ref{fid_score_30}. We calculate
the mean fidelities averaged over these states,
$\bar{\bar{f}}$, with and without discard, which are then
plotted against $\epsilon$ in Figure~\ref{fid_score_30}(a).
This graph shows an improvement as we increase the amount of
discard. We also present our results through a
score plot, where we compute the number of states out of 2000
starting states for which our scheme outperforms the
projective measurement scheme.  The score plot is described
in Figure~\ref{fid_score_30}(b).  Interestingly, this number
crosses the 50\% mark for a threshold value of the discard
parameter.

When a study of mean fidelity, averaged over 2000
states, $\bar{\bar{f}}$ vs $\epsilon$ was done, it turns out
that although on the average the performance of projective
measurements is better, if ambiguous meter readings are
discarded, then the number of states for which our
tomography scheme is successful, goes up. In fact, number of
successes out of 2000 for the discard parameter values of
0, 0.2, 0.4, 0.6 and 0.8 
are 923, 973, 1023, 1051 and 1071,
respectively.
This we think is a clear evidence that our
scheme has the potential of unearthing more information than
projective measurements under certain circumstances.
In particular,
if we are given $30$ copies of a unknown state of
a qubit, our scheme will be a better choice for carrying out
state tomography.

We now turn to testing our scheme with increasing ensemble size.
We repeat the simulation in exactly the same way for the
case of ensemble size $60$.
The results are presented in a similar way in
Figure~\ref{fid_score_60}.
Increasing the ensemble size clearly reduces the
efficacy of our scheme as compared to projective measurements.
The score plot show that our scheme outperforms the
projective measurement scheme for ensemble size of $60$
for lesser number of states and the number is less than
$50\%$. Therefore we conclude that our scheme is preferable
only when we have a small ensemble size. We would
like to clarify that this not due to statistical
fluctuations as we have taken the average over a large number of
runs even when the ensemble size is small. For a more extensive
study of state estimation by this scheme see~\cite{Qsize}.

\section{Concluding remarks}
We have presented a scheme for state estimation
based on weak measurements. The weak or unsharp
measurements that we have used are those where the
apparatus system coupling is weak. In this regime,
although the information obtained from the system
is limited, the corresponding disturbance caused to
the state is also small. Thus the possibility of
re-using the states becomes available. We show that
for small ensemble sizes, the weak measurement based
scheme can outperform the projective measurement
based scheme. This opens up new possibilities for
extracting information from quantum systems.

We have recently extended these results to the
domain of continuous variable systems with one degree
of freedom. We explore how a weak measurement
tomography scheme can be used to estimate the Gaussian
state of such systems. We have interesting and
encouraging preliminary findings in this context,  
which will be presented in a forthcoming publication.

\end{document}